# DIRECT FOCUSING ERROR CORRECTION WITH RING-WIDE TBT BEAM POSITION DATA

M.J. Yang[‡], Fermilab[†], Batavia, IL 60510, USA

*Abstract*

Turn-By-Turn (TBT) betatron oscillation data is a very powerful tool in studying machine optics. Hundreds and thousands of turns of free oscillations are taken in just few tens of milliseconds. With beam covering all positions and angles at every location TBT data can be used to diagnose focusing errors almost instantly. This paper describes a new approach that observes focusing error collectively over all available TBT data to find the optimized quadrupole strength, one location at a time. Example will be shown and other issues will be discussed.

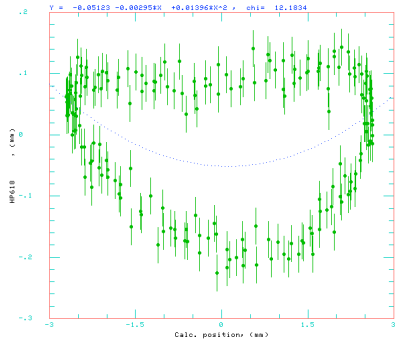

Figure 1. Horizontal position deviations at HP618 are plotted against model's expected positions for 200 consecutive turns.

## INTRODUCTION

TBT data for lattice function measurement has been presented before [1] and has been used at Fermilab Main Injector [2] for many years. Although most of the measured results were within 10-15% of design lattice function some deviated more substantially. One example would be the measurement made at 150 GeV flat-top energy. The procedure to be outlined in this paper was developed while examining this TBT data for quadrupole errors.

It was first noticed that TBT position deviations, from expected positions calculated by model, tend to form oddly shaped contours when plotted against the expected positions. Figure 1 shows one such contour as example, where data from HP618 is plotted. These contours turned out to be signatures of focusing error. By adjusting quadrupole magnet strength upstream it is possible to collapse the contour down to a band-shaped distribution.

To use this signature to guide the search for quadrupole errors it is necessary to treat the machine as a beamline, not a circular ring, and watch the development of position deviation contour one location at a time. The model will need to be adjusted to match the data wherever quadrupole error is determined. The methodology will be discussed in detail and some result will be shown.

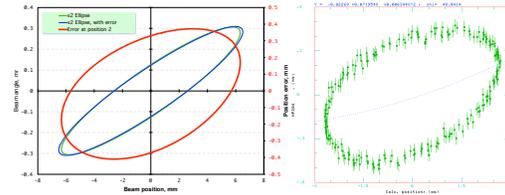

Figure 2. Simulated position deviations due to quadrupole error, matching that of HP204. Actually, a tiny amount of sextupole had to be introduced to make upper-right end look a little bit pointed, as seen in data.

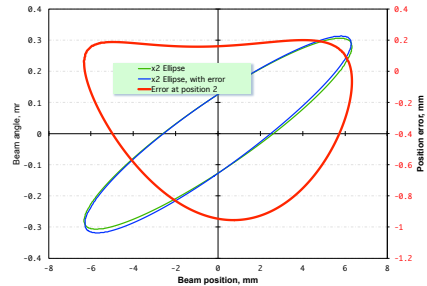

Figure 3. Simulation of position deviations at HP618, as shown in Figure 1. The ellipse transported without error is plotted in green and ellipse with error in blue. Both reference to the vertical axis on the left. The red contour shows simulated position deviation vs. expected position and reference to the vertical axis on the right.

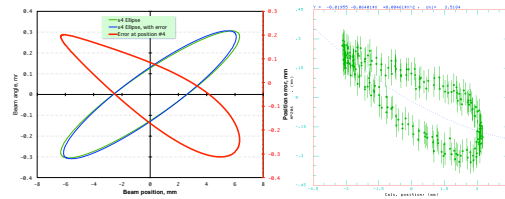

Figure 4. Teardrop shaped contour at HP206.

## SIMULATING POSITION DEVIATIONS

To get an understanding of what it takes to produce position deviation contours as observed a 2-cell FODO lattice model was constructed on an Excel spreadsheet. A phase space ellipse was transported twice, once with perfect optics and the second time with quadrupole error and also with sextupole field.

With quadrupole errors used in simulation symmetric ellipse-shaped contour, such as the one shown in Figure 2, is observed. With addition of sextupole the contours can take up many different forms. Figure 3 shows a simulation of data from HP618, as shown in Figure 1. Other examples, along with data being simulated, are shown in Figure 4, 5, 6, and 7. It became clear that sextuple effect dominates the shaping of contours. Those different sets of parameters used to match data contours at various locations can be rationalized as matching the accumulated effect from upstream quadrupole errors and sextupoles.

Being able to simulate these data contours is important.

It established a necessary condition, i.e. quadrupole is effective in correcting observed position deviations.

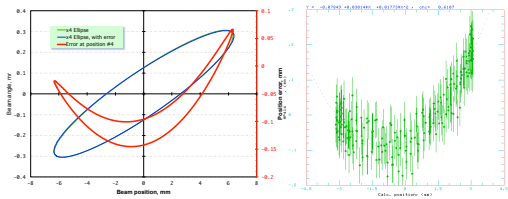

Figure 5.   V-band shaped contour at HP414.

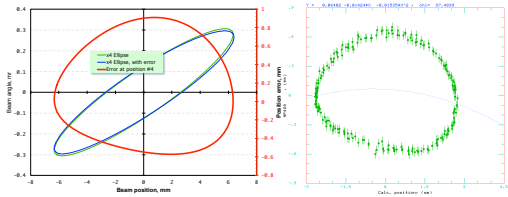

Figure 6.   Tumbled-rock shaped contour at HP428.

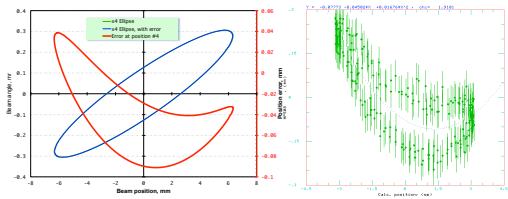

Figure 7.   Banana shaped contour at HP508.

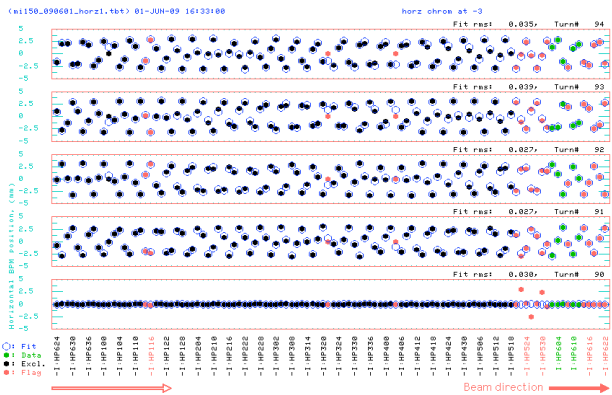

Figure 8.   TBT initial orbit parameters were fitted using only five data points, plotted in green dots. With VP601 as the start of beamline HP602 is the first horizontal plane BPM, the first green dot. For each turn data progression is from left to right and wraps around on the left. Bottom plot, turn #90, showed that beam was kicked at MI 520, five BPMs before start of next turn.

## DATA

Fermilab Main Injector has 106 BPMs in the horizontal plane and 109 in the vertical plane, providing up to 2048 turns of TBT data each. Extraction kicker at MI52 was used to excite free betatron oscillation in the horizontal plane. There are 200 turns of horizontal plane TBT data, taken at 150 GeV flat-top, used in the analysis presented here. Figure 8 shows five turns of full ring data, from turn #90 to #94. BPM data used in the analysis has been corrected with gain calibrations derived from TBT data analysis [3].

For beamline starting at VP601 five horizontal plane BPMs from locations 602, 604, 606, 610, and 612 were used to fit for TBT initial orbit parameters. Because of its higher gain response BPM HP608 was excluded in the fit. Though gain correction has definitely brought it back in line with others it is still on the watch list.

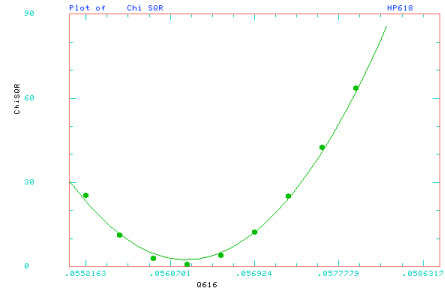

Figure 9.   Strength of Q616 was scanned and corresponding chi-square values, from fitting position deviations at HP618, are plotted. The horizontal plot axis is in KG/M/Amp with default strength right at the middle. The location of minimum can be calculated from the second order polynomial fit as shown.

## QUADRUPOLE STRENGTH SCAN

Position deviations at HP618, and shown Figure 1, is used as example to demonstrate the quadrupole strength scanning procedure. Quadrupole Q616, at 107° in phase advance ahead of 618 location, is the natural choice for the scan. Also shown in the figure is the result of a second order polynomial fit to the plotted position deviations. The fitted chi-square serves to characterize the open-ness of a contour and is the value to be minimized.

Plotted on Figure 9 are chi-square values as a function of Q616 strengths used in the scan, covering a range of ±3.0% from the default quadrupole strength. By setting new quadrupole strength to where chi-square minimum is, -1.1% from default, the resulted position deviation contour collapsed into a V-shaped band seen in Figure 10. Now, it is easy to see why a second order polynomial fit is needed, i.e. to deal with the second order effect that model is not accounting for.

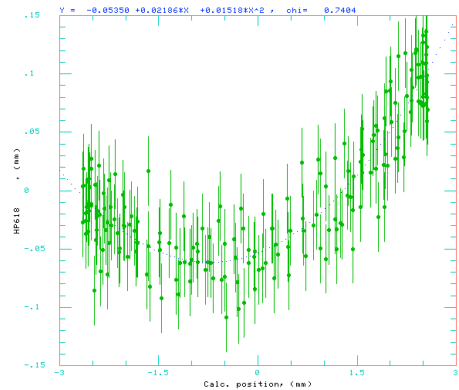

Figure 10.   HP618 error after optimized strength for quadrupole Q616 was used in the model calculation.

## WORKING ON BEAMLINE

*Fitting for initial orbit parameters*

Since all data turns are independent initial orbit parame-ters need to be extracted from data individually. It is important that start of beamline is chosen where accurate initial orbit parameters can be extracted from available data. For this analysis five BPMs were used and they are shown as green dots in turn data plots of Figure 8. The corresponding RMS of fit, shown above each turn plot, is consistent with the RMS noise of BPMs used. By nature TBT turn data can be re-segmented into individual passes of data, for beamline beginning at any given location.

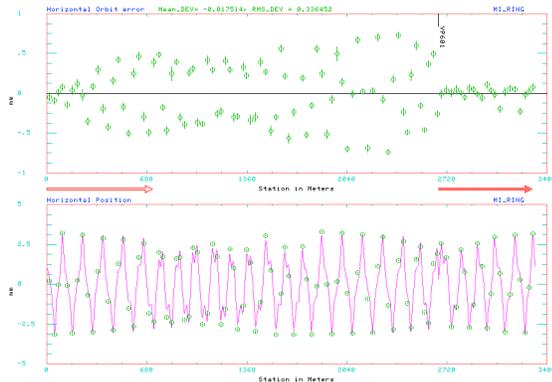

Figure 11. Example of single turn orbit data. Bottom plot shows BPM data in green circles and the matching model calculation in magenta trace. Top plot shows position deviations between data and model calculation.

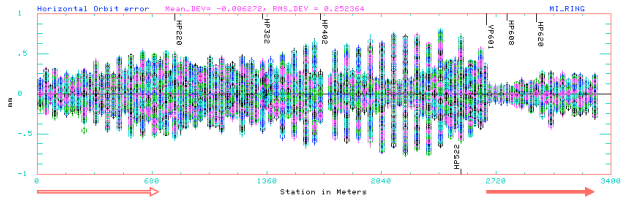

Figure 12. Overlaid plot of horizontal plane position deviations from all 200 turns used in the analysis.

### Expected orbit from Beamline Model

For each turn the model propagates initial orbit parameters down the beamline to every BPM location. Figure 11 shows the comparison between data and model calculation for turn #107. The beamline starts from VP601, as indicated by solid red arrow, moves to the right, and wraps around on the very left. The deviations are small at the beginning and grow larger as beamline continues. Figure 12 gives an overall view with overlaid plot of position deviations from all 200 turns.

### Scanning position deviation data

Starting from beginning of beamline position deviations were examined one location at a time, and quadrupole scans were performed in the way described above. The improved matching between data and model calculation, using modified quadrupole strengths, can be seen in Figure 13 and 14. Same data, turn #107, is plotted in Figure 13 for comparison with Figure 11. The deviations clearly are much smaller. Substantial improvement can be seen when comparing Figure 14 to Figure 12, where deviations from all 200 turns are plotted.

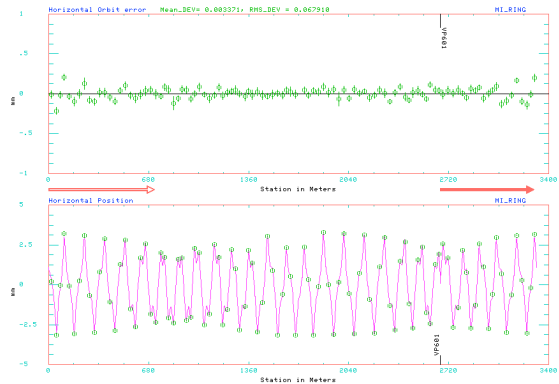

Figure 13. Same turn #107 data is plotted. The newly found quadrupole strengths are used for model calculation. The match between calculation and data is clearly much better.

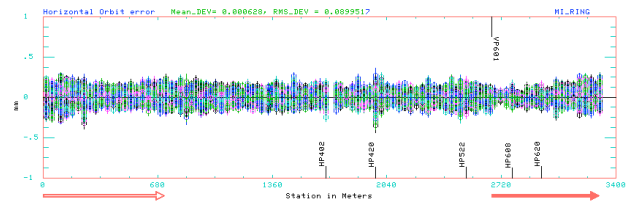

Figure 14. Overlaid plot of 200 turns of position deviations with modified quadrupole strength used in model calculation.

### The errors

It is important that fitted initial orbit parameters do not inject non-existent error that will require un-warranted corrections later. A complimentary analysis with beamline starting at a different location would provide an important second diagnosis.

Several factors may be contributing to the residual deviations seen in Figure 14, for example, poorly executed scan that leads to bad correction, a few percent calibration error of BPM reading, and the still un-accounted for sextupole or even higher order effects.

## CONCLUSION

The procedure presented clearly has helped to reduce overall deviations significantly, with relative ease. Sextupoles, being a permanent feature of the ring, will need to be incorporated into the model. While cumulative effect from all sextupoles around the ring may be negligible on turn-to-turn basis it is not so in this transfer line analysis.

It should be noted that this procedure is not limited to looking for quadrupole errors. By modifying the target of minimization it could in principle be used to look for skew quadrupole errors and sextupole errors as well.